\documentclass[a4paper]{VisionStyle}
\usepackage{epsfig}

\begin{document}

\title{Chandra Observations of the Gravitational Potential Structure in
Abell 1060}

\author{T.\,Furusho\inst{1} \and N.Y.\,Yamasaki\inst{2} \and 
  T.\,Ohashi\inst{2} } 

\institute{Laboratory for High Energy Astrophysics, NASA/GSFC, Code
662, Greenbelt, MD 20771, USA \and Department of Physics, Tokyo
Metropolitan University, 1-1 Minami-Ohsawa, Hachioji, Tokyo 192-0397,
Japan}

\maketitle 

\begin{abstract}

We present results from a Chandra observation of Abell 1060, the
nearby isothermal cluster of galaxies. The ACIS-I image shows that the
central cusp-like structure, previously seen from PSPC, is caused
mainly by an emission from the central elliptical galaxy NGC~3311\@.
We confirmed that the central region is remarkably isothermal with a
temperature of 3.2 keV, based on the spectral fits for fine pixels. An
extended region in the northeast of NGC 3311 indicates high iron
abundance. The surface brightness profile excluding the central
galaxy, NGC 3311, can be fitted only by a double $\beta$ model, with
core radii about 40 and 140 kpc, out to a radius of about
200 kpc. This suggests that the gravitational potential
in Abell 1060 consists of 2 components.

\end{abstract}

\section{Introduction}

The gravitational potential structures of clusters have been probed mainly by X-ray
observations of the intracluster medium (ICM)\@. ICM distributions are
empirically represented well by an isothermal $\beta$ model, which approximates 
the gas density profile given by the King-type potential with a
flat density core under the assumption of hydrostatic equilibrium.  On the other hand, recent numerical simulations
(e.g.\ \cite{tfurusho-B3:nfw96}) involving a large number of test
particles indicate that the potential profile takes a universal form
with a cusp structure in the central $\sim 50$ kpc region.  For an
observational determination of the gravitational potential structure,
an X-ray observation with high angular resolution is the most powerful way
because it enables us to map out the gas density and temperature right
into the center of the cluster.

Abell 1060 (hereafter A1060) is an X-ray bright cluster of galaxies at a
redshift of $z=0.0114$.  There are two giant elliptical galaxies in the
center, NGC 3311 as a central cD galaxy and NGC 3309. Previous X-ray
observations with ROSAT and ASCA have shown that A1060 has circularly
symmetric surface brightness, an average temperature of 3.3 keV, and a
constant abundance of 0.3 solar (\cite{tfurusho-B3:tam96},
\cite{tfurusho-B3:tam00}, \cite{tfurusho-B3:fur01}). Any significant
feature of a cool component or cooling flow is almost absent in the
spectrum of the central region. The cluster optical morphology showed
that A1060 is remarkably isolated in redshift space
(\cite{tfurusho-B3:ric82}).  These unique properties allow us to look
into the gravitational potential structure which is almost free from
external effects such as subcluster mergers.

In this paper, we use $H_0=75$ km s$^{-1}$ Mpc$^{-1}$ and $q_0 = 0.5$,
which indicates $1''=0.217$ kpc and $1'=13$ kpc at the cluster.  The
solar number abundance of Fe relative to H is taken as $4.68 \times
10^{-5}$ (Anders and Grevesse\ 1989).

\section{Observation and analysis}
\label{tfurusho-B3_sec:tit}

The Chandra observation was performed on 2001 June 4 with ACIS-I,
consisting of CCD chips I0123, and S2, for an exposure time of 32
ks. The data were obtained with the Very Faint mode. The cluster
center was focused at the center of I3 chip. Except for 1--2 ks at the
end of the observation, the count rate in the energy band 0.3--10 keV
stayed almost constant at 4.1 counts s$^{-1}$ for the I3 chip, and 2.9
counts s$^{-1}$ for the I012 chips, respectively.  Since the
contribution from the particle background was relatively high in I012
chips, which do not cover the cluster center, we excluded all the time
intervals when the count rates of these chips exceeded 3.2 counts
s$^{-1}$. This screening resulted in a useful exposure time of 29.8
ks. The background image and spectrum were subtracted based on the
blank sky data prepared by M. Markevitch
(http://asc.harvard.edu/cal/). The average background count rate was
0.27 count s$^{-1}$ chip$^{-1}$ in the 0.3--10 keV band, which was less
than the cluster flux by a factor of about 10 and 4 for the I3 and
I012 chips, respectively.

\section{Results}
\label{tfurusho-B3_sec:cmd}

\subsection{The central region}
\label{tfurusho-B3_sec:ex1}

\begin{figure}[ht]
  \begin{center}
    \epsfig{file=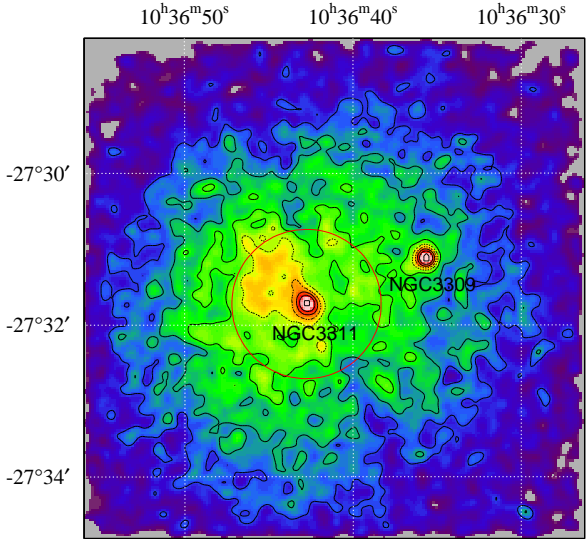, width=8.5cm}
  \end{center}
\caption{The central ACIS-I3 image of A1060 in 0.5--10 keV band. The red circle shows
roughly the optical size with a radius of $1'=13$ kpc.}
\label{tfurusho-B3_fig:fig1}
\end{figure}

The X-ray image of the central $6.\!'6\times6.\!'6$ region of the
cluster taken with the I3 chip is shown in
Figure~\ref{tfurusho-B3_fig:fig1}. Contours are smoothed by a Gaussian
function with $\sigma=4''$, and corrections for background and exposure
time are performed. The central elliptical galaxies, NGC 3311 and NGC
3309, are clearly resolved, and a diffuse excess emission in the
northeast of NGC 3311 is also revealed.  The X-ray properties of the two
elliptical galaxies are described in \cite*{tfurusho-B3:yam02}. Here, we
report on the northeast emission of NGC 3311.

\begin{figure}[ht]
  \begin{center}
    \epsfig{file=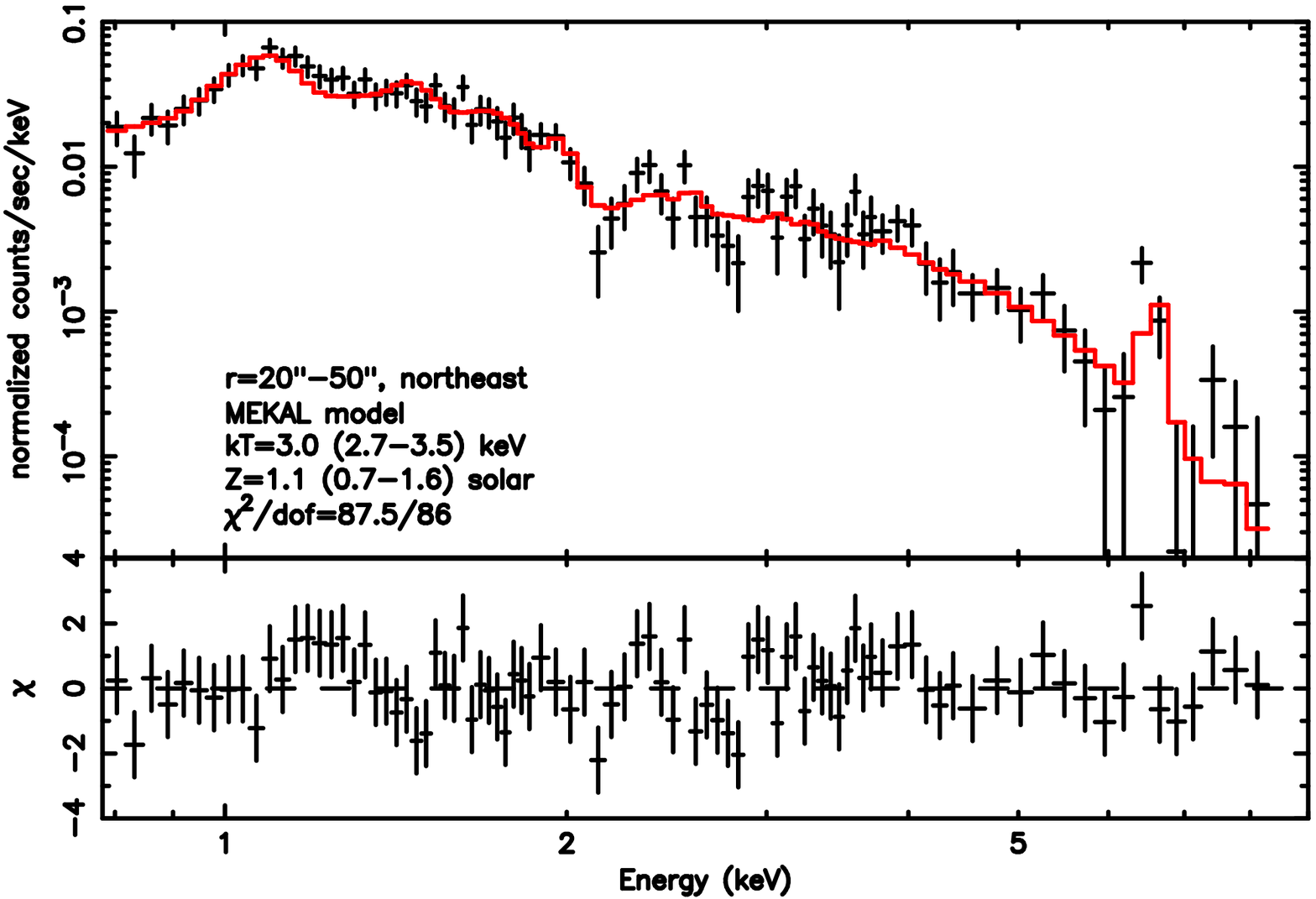, width=7.2cm}
    \epsfig{file=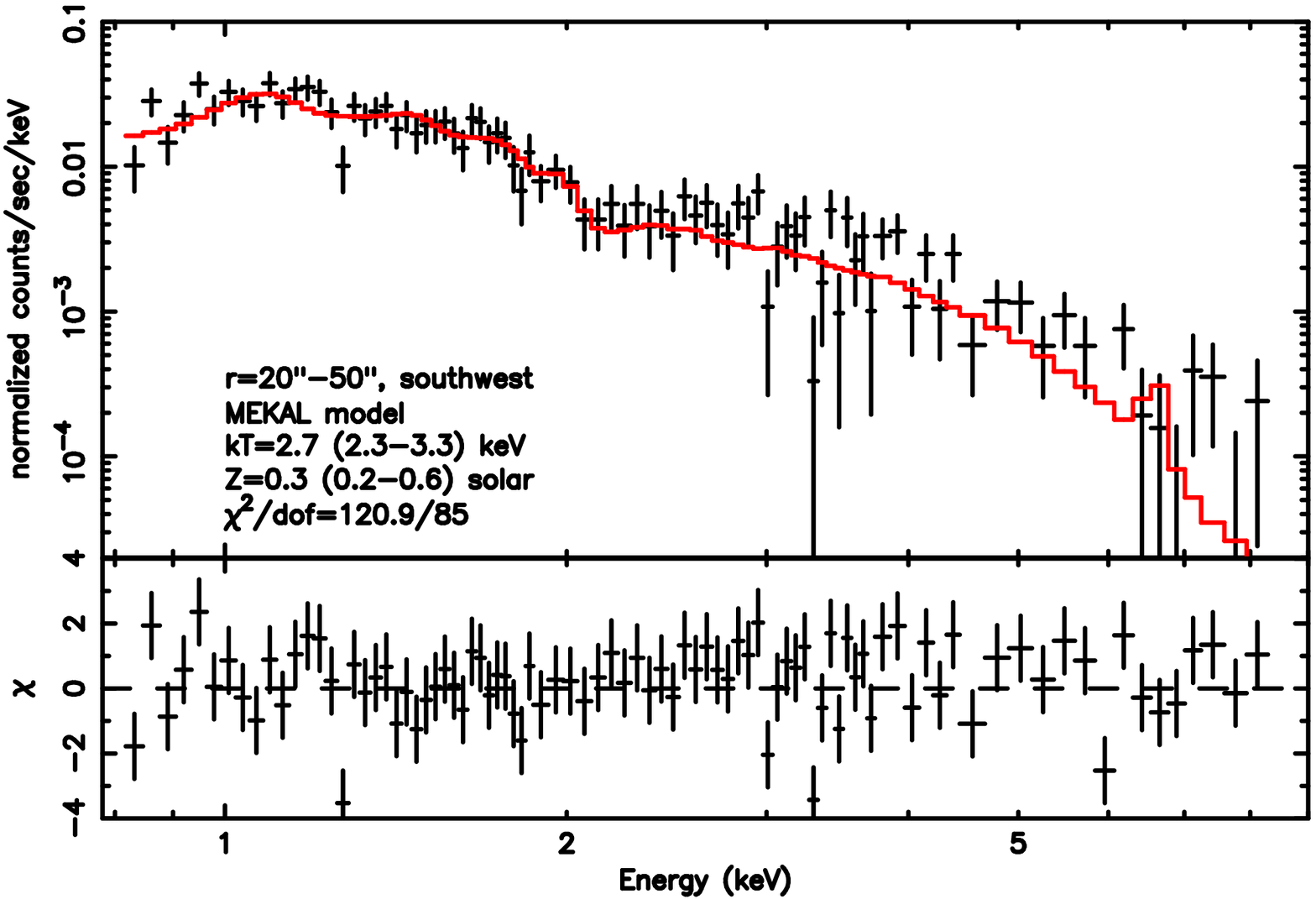, width=7.2cm}
  \end{center}
\caption{ Pulse-height spectra and best-fit absorbed MEKAL model for the
northeast (top) and southwest (bottom) annular regions of the NGC 3311 with a
radius of $20''-50''$.}
\label{tfurusho-B3_fig:fig2}
\end{figure}

Figure~\ref{tfurusho-B3_fig:fig2} shows the spectrum for
the northeast annulus with a radius range of $20''-50''$, compared
with that in the opposite southwest region. Both spectra indicate a
similar temperature of about 3 keV, which is consistent with the ICM
level. However, their Fe-line features are markedly different. The
best-fit abundance is as high as 1.0 solar for the northeast region,
much higher than the 0.3 solar observed both in the southwest region
and in the general ICM of A1060. The excess X-ray luminosity in this
region is estimated to be several $\times10^{40}$ erg s$^{-1}$, which is
comparable to the emission of NGC 3311 itself. We note that the region
with the anomalous abundance is located within the isophotal radius of
NGC 3311 (21 kpc, \cite{tfurusho-B3:vas91}).

\subsection{Temperature map}
\label{tfurusho-B3_sec:ex2}

The temperature map of A1060 derived from hardness ratio analysis on the
ASCA GIS data showed that the ICM is almost isothermal with an average
temperature of 3.3 keV in a scale of $5'$ over the whole
cluster(\cite{tfurusho-B3:fur01}). The angular resolution of Chandra
allows us to perform spectral fits in much smaller scales, since the
data are not affected by stray light and energy-dependent PSF effects.
To produce the temperature map with sufficient statistics, we divided
the image into 9 annulus regions, which are further divided into 2--4
azimuthal sections.  Point source candidates and NGC 3309 were masked
out with a circle of $20''$ radius.

The spectra can be represented by an absorbed MEKAL model, however, all
the data showed absorption features that were considerably in excess of
the Galactic value of $N_{\rm H}=6\times10^{20}$ cm$^{-2}$. Since no
significant absorption was seen in the PSPC data, this is likely to be
caused by an incorrect response function in the low energy band. To
derive the temperature, we tried two methods. One was to fit the spectra
in the energy band 0.8--8 keV by adjusting $N_{\rm H}$ as a free
parameter, and the other was the fit in 2--8 keV with fixed $N_{\rm H}$
at the Galactic value. The temperature difference between the two fits
is very small and 0.5 keV at most. The typical statistical errors are
0.3 keV at the 90\% confidence. 

The resultant temperature map based on the free $N_{\rm H}$ fit is shown
in Figure~\ref{tfurusho-B3_fig:fig3}. It is shown that the temperature
distribution in the ICM is very uniform with a temperature around 3.2
keV. The most inner regions at the south of NGC 3311 give slightly lower
temperatures of $\sim 2.5$ keV. This small temperature drop at the
center is considered to be caused by a contamination of NGC 3311
component, not a cooling flow.

\begin{figure}[ht]
  \begin{center}
  \end{center}
\caption{ Color-coded temperature map of the ACIS-I. The contours
indicate the X-ray image smoothed by a Gaussian function with $\sigma=8''$.}
\label{tfurusho-B3_fig:fig3}
\end{figure}

\subsection{Surface brightness and mass profiles}
\label{tfurusho-B3_sec:ex3}

\begin{figure}[ht]
  \begin{center}
(a)\hspace*{4.2cm}(b)
\hspace*{-6mm}\epsfig{file=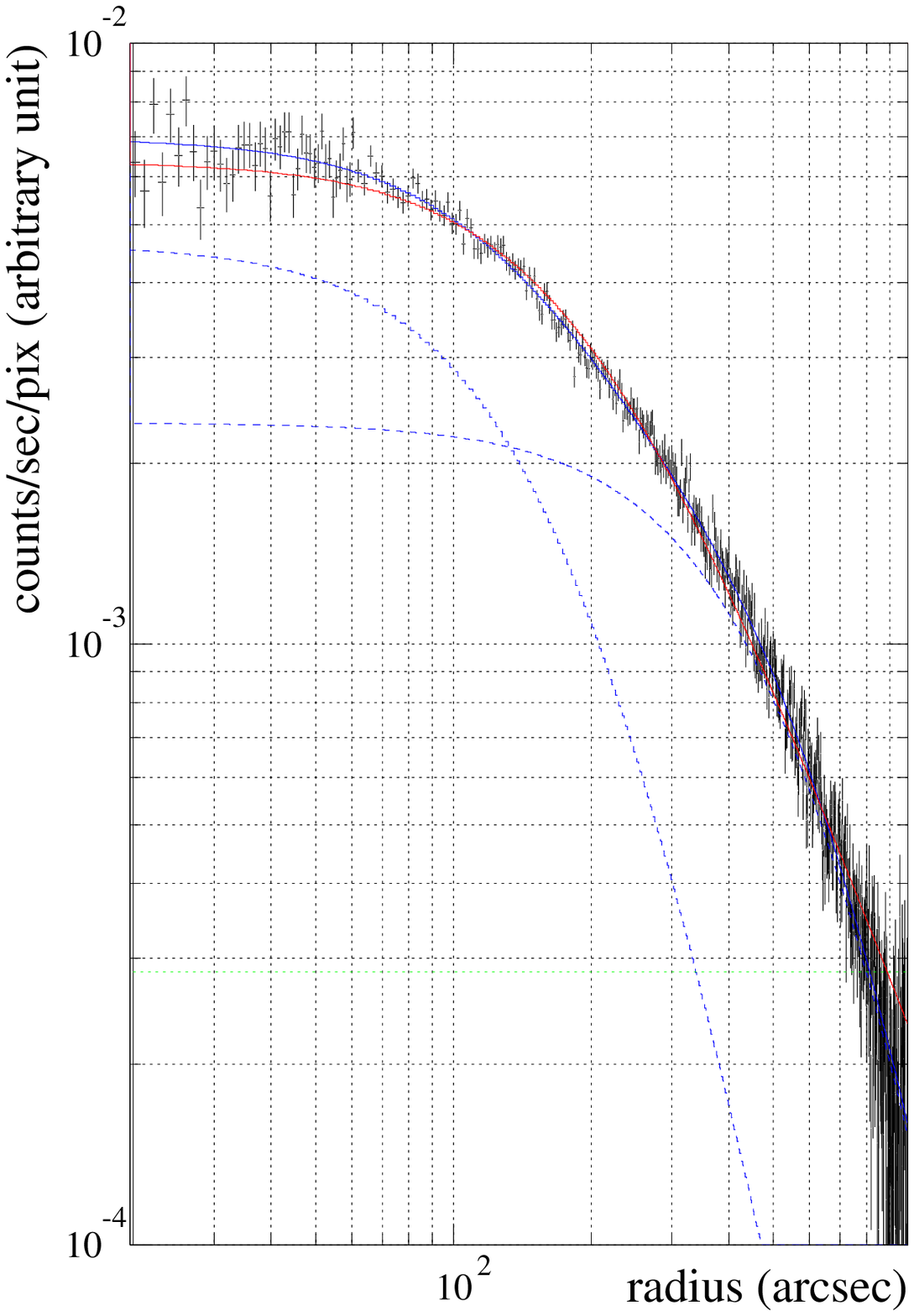, width=4.6cm}
\hspace*{-3mm}\epsfig{file=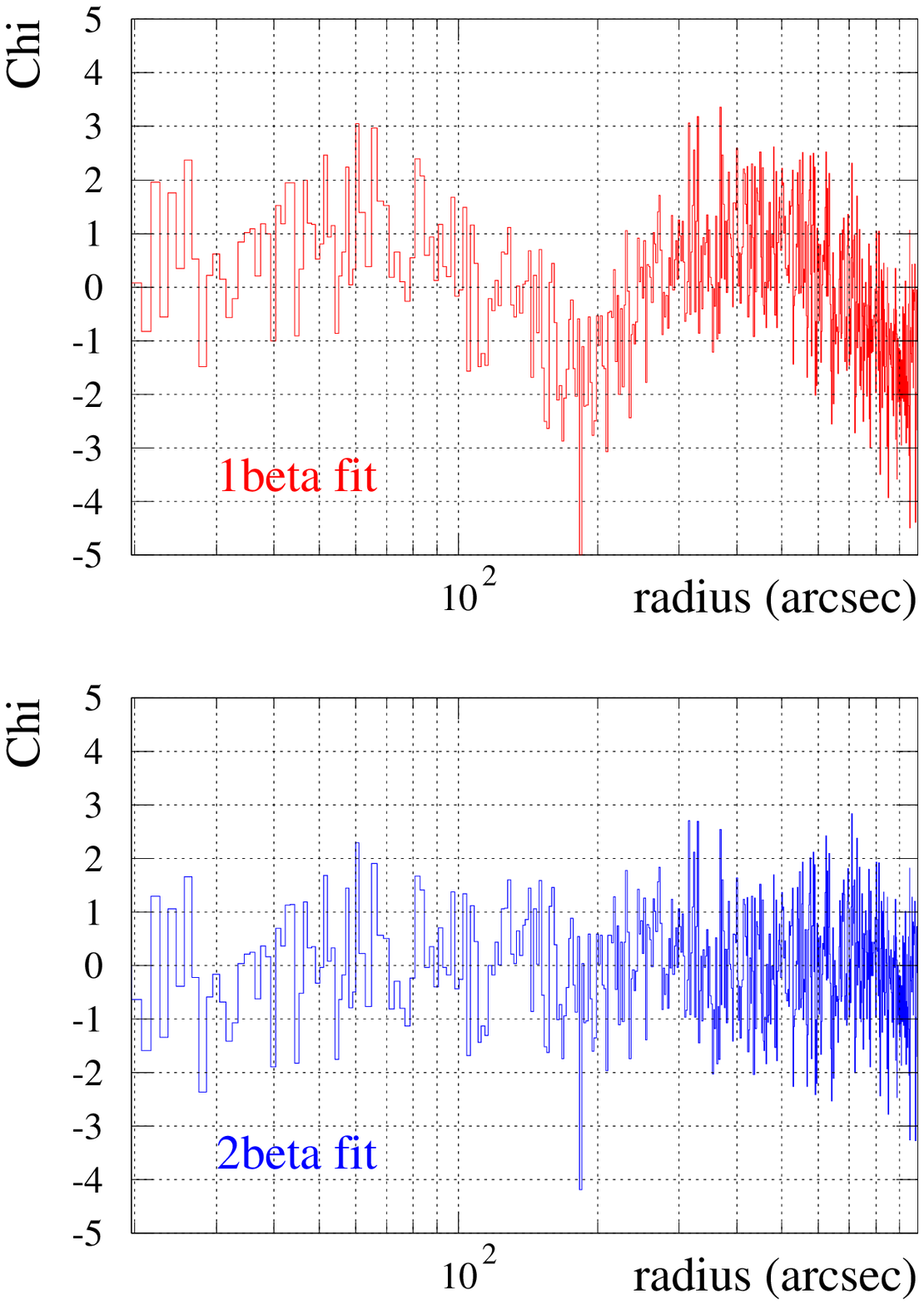, width=4.6cm}
  \end{center}
\caption{(a) X-ray surface brightness profile and the best-fit single (red)
and double (blue) models with the subtracted background level (green
dotted line). (b) The residual from the best-fit single (top)
and double (bottom) $\beta$ models. }
\label{tfurusho-B3_fig:fig4}
\end{figure}

 Figure~\ref{tfurusho-B3_fig:fig4} shows the surface brightness profile
in the 0.5--10 keV energy band after subtraction of the background, point
sources, and the diffuse emission in the northeast of NGC 3311. The data are
fitted with single $\beta$ and double $\beta$ models for the region of $20''<r<15'$.  The single
$\beta$ fit gives $r_{\rm c}=46.0\pm1.2$ kpc and $\beta=0.51\pm0.1$,
which are both slightly different from the PSPC results of $r_{\rm
c}=50.7\pm1.3$ kpc and $\beta=0.54\pm0.1$
(\cite{tfurusho-B3:tam00}). This may be due to the changes of covered
energy band and the observed area, for which the PSPC parameters were
0.5--2 keV and $r<60'$.  The profile is, however, not simply described
by the single $\beta$ model indicating $\chi^2/dof=823.3/465$. The NFW
model gives $\chi^2/dof=694.4/464$, which is not a good fit either. If
we fit the profile by changing the outer cut-off radius in $r=3'-15'$,
the best-fit core radius and $\beta$ increase with the cut-off
radius. Based on this feature, we tried a double $\beta$ model, which
could fit the data well showing $\chi^2/dof=532.8/462$. The best-fit
core radii were obtained to be 41.5 and 142.7 kpc with $\beta=0.82$ and 0.94, respectively. These values are
close to the 2 peaks (40 and 173 kpc) in the distribution of
core radius for 79 distant ($z>0.1$) clusters, recently studied by
\cite*{tfurusho-B3:ota01}.

\begin{figure}[ht]
  \begin{center}
    \epsfig{file=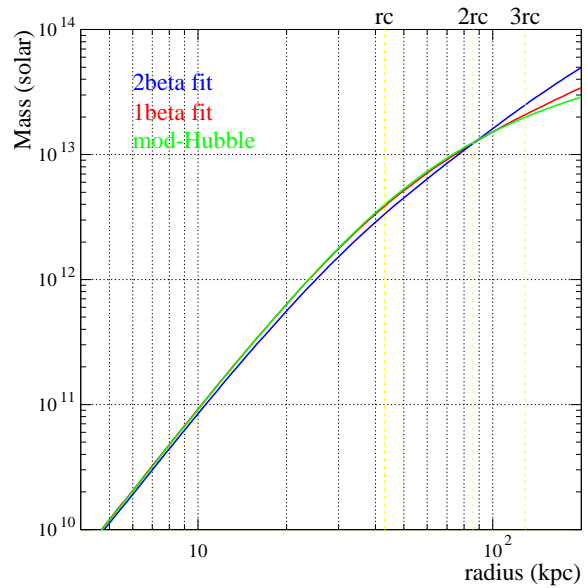, width=8.5cm}
  \end{center}
\caption{ Total mass profiles from the single (red) and double (blue)
$\beta$ models, and modified Hubble model (green). The $r_{\rm c}$ at
the top indicates 44 kpc that is the best-fit core radius fitted with a
single $\beta$ model.}
\label{tfurusho-B3_fig:fig5}
\end{figure}

Because of the symmetric morphology and the isothermality seen in the
temperature map, it is a reasonable assumption that this cluster is in
a hydrostatic equilibrium.  Under this assumption, we can estimate the
total gravitating mass profile for the single and double $\beta$
models, as shown in Figure~\ref{tfurusho-B3_fig:fig5}.  The mass
profile assuming the modified Hubble law is also shown for comparison.
There is some difference between the mass distributions for the single
and double $\beta$ models, and the modified Hubble law which is the
approximation of the King potential gives a similar profile with the
single $\beta$ case.

\section{Discussion}
\label{tfurusho-B3_sec:dis}

The excess X-ray emission extending to the northeast of the central
galaxy NGC 3311 indicated an unusually high metal abundance with a
temperature of about 3 keV.  One possibility is that the metal rich gas
was ejected from NGC 3311 by a galactic wind, which may be heated by a
collision with the surrounding ICM\@. The excess region is, however,
located inside the galaxy in the optical band, and the size and mass of
NGC 3311 are smaller than those of typical cD galaxies. Also, NGC 3311
shows weak radio emission with no jet-like feature around the core
(\cite{tfurusho-B3:lin85}). These features rather suggest that the metal
rich gas could be stripped off from NGC 3311. In this case, this galaxy
is passing the core region and its location at the center may be a
chance coincidence. The very compact X-ray halo, like the one in NGC
3309, also suggests this possibility.

We have confirmed that the core region of A1060 has a smooth and
circularly symmetric X-ray structure. The gas within $r<15'$ is fairly
isothermal with $kT\sim3.2$ keV on $\sim 1'$ scales. Optical
observations detected a huge void of about 50 Mpc extent in the front
and back of the A1060 cluster. These features jointly suggest that the
cluster has never experienced major mergers in the recent $\sim1$ Gyr. 
Also, the central gas density of $3\times10^{-3}\ {\rm cm}^{-3}$
is lower than those in other cD clusters,
resulting in a very long cooling time. Therefore, the formation
process of A1060 may be going slowly and the cluster stays in a
relatively young stage, which is also consistent with the low galaxy
density in the surrounding region.

The surface brightness profile shows that the central cusp-like
structure, reported by \cite*{tfurusho-B3:tam00}, is not due to the
dark halo potential, rather it is a combination of the central galaxy
NGC 3311 and the ICM\@. Even after subtraction of the central galaxy,
the profile requires at least two components. The deviation from the
single component model cannot be explained by an error in the
approximation of the isothermal potential, and it seems that the
gravitational potential itself has a double structure. Since A1060 is
considered to be young and evolving, its potential may correspond to a
structure taking place in an early phase of the cluster formation.  If
this is the case, the gravitational potential may gradually change its
shape from a double to a single structure while the intracluster gas
falls onto the center, as proposed by \cite*{tfurusho-B3:ota01}.

\begin{acknowledgements}
The authors thank to Dr.\ S. Sasaki for useful comments.
T.F. is supported by the Japan Society for the Promotion of Science
Postdoctoral Fellowships for Research Abroad.
\end{acknowledgements}

\end{document}